\documentclass[11pt,a4paper]{article}
\usepackage{jcappub}

\title{Non--linear weak lensing forecasts}

\author[a,b]{Luciano Casarini,}
\author[a,b]{Giuseppe La Vacca,}
\author[c]{Luca Amendola,}
\author[a,b]{Silvio A. Bonometto}
\author[d]{\& Andrea V. Macci\`o}

\affiliation[a]{Department of Physics G.~Occhialini, Milano--Bicocca
University\\ Piazza della Scienza 3, 20126 Milano, Italy}
\affiliation[b]{I.N.F.N., Sezione di Milano--Bicocca \\ 
Piazza della Scienza 3, 20126 Milano, Italy} 
\affiliation[c]{Institute of Theoretical Physics \\ 
Philosophenweg 16, 69120 Heidelberg, Germany} 

\affiliation[d]{Max-Planck-Institut f\"ur Astronomie \\ 
K\"onigstuhl 17, 69117 Heidelberg, Germany}

\emailAdd{luciano.casarini@mib.infn.it}
\emailAdd{giuseppe.lavacca@mib.infn.it}
\emailAdd{l.amendola@thphys.uni-heidelberg.de}
\emailAdd{bonometto@oats.inaf.it}
\emailAdd{maccio@mpia.de}

\abstract{We investigate the impact of non-linear corrections on dark
energy parameter estimation from weak lensing probes. We find that
using {\sc halofit} expressions, suited to $\Lambda$CDM models,
implies substantial discrepancies with respect to results directly
obtained from $N$--body simulations, when $w(z) \neq
-1$. Discrepancies appear strong when using models with $w'(z=0)>0$,
as fiducial models; they are however significant even in the
neighborhood of $\Lambda$CDM, where neglecting the degrees of freedom
associated with the DE state equation can lead to a misestimate of the
matter density parameter $\Omega_m$.}

\keywords{cosmology: theory, dark matter, gravitation; methods:
numerical, N--body simulations.}

\begin{document}

\maketitle

\section{Introduction}
\label{sec:intro}

The evidence in favor of dark energy (DE), a smooth component with
largely negative state parameter $w \sim -1$, is more than a decade
old. The original claim based on the acceleration in the expansion
rate, deduced from the Hubble diagram of supernovae Ia \cite{bib1},
has been confirmed and strengthened by other probes, in particular
cosmic microwave background (CMB) spectra (see, e.g., \cite{wmap7}) and
large scale structure (LSS) data \cite{bib3}. The nature of DE is
however one of the major puzzles of today's physics and
astrophysics. In order to investigate it, more data constraining the
DE state equation are sought.

Data based on the CMB spectra $C_\ell$, however, only scarcely
constrain the equation of state parameter of the Dark Energy component
$w(z)$. In fact, although the integrated Sachs \& Wolfe (ISW) effect
depends on $w(z)$, it mostly affects low--$\ell$ $C_\ell$ which are
also affected by the $z$--dependence of the optical depth $\tau$.
Furthermore, ISW is an integral effect, and low--$\ell$ spectral
components are subject to a large cosmic variance.

Putting together CMB and LSS data is already more effective, as this
tests the fluctuation distribution and amplitude at two different $z$
values. Clearly, estimating the fluctuation spectrum $P(k,z) = \langle
\left| \delta_m(k,z) \right|^2 \rangle$ at several $z$ values, so to
work out the growth rate
\begin{equation} 
G(z,k)=\delta_{m}(k,z)/\delta_{m}(k,0)~,
\end{equation}
would open a real window on $w(z)$ and, therefore, on DE nature. So
far $G$ is only very weakly constrained (see, e.g., \cite{bib4}), but
a formidable array of surveys are being performed or planned to
measure it through weak lensing (WL), high--$z$ power spectra and
cluster counts \cite{bib5}.

This paper will deal with WL probes (see, e.g., \cite{bib6} for a
thorough review). The power of this tool is illustrated by the fact
that, to fully exploit next generation WL surveys, we need predictions
on non--linear power spectra accurate up to $\sim1\%$ \cite{bib7}.

Such precision goes beyond the claimed $\pm 3\, \%$ accuracy of the
popular {\sc halofit} expressions \cite{hlf} based on the halo model
of structure formation and calibrated using numerical simulations of
$\Lambda$CDM models; they yield a map of linear $\kappa$'s onto
non--linear $k$'s, so that $P_{nl}(k) = P_{lin}(\kappa)$. Leaving
apart the need of a generic improvement of {\sc halofit} 
accuracy \cite{hilbert},
previous work \cite{mcdo} already showed that, using {\sc halofit}
expressions for non--$\Lambda$CDM models, requires suitable
corrections. In spite of that, the {\sc halofit} map has been often
used to obtain the spectra of models with non--constant DE state
parameter $w(z)$, that we shall denominate {\it dynamical DE} (dDE)
models, herebelow. This procedure was dictated by the lack of
appropriate extensions of {\sc halofit} to non--$\Lambda$CDM
cosmologies.

The aim of this paper is to quantify the effects of using non--linear
dDE spectra obtained with the {\sc halofit} map, when the nature of DE
is investigated through WL surveys as Euclid \cite{euclid2} or COSMOS
\cite{cosmos}.

To this aim we focus on dDE models with
\begin{equation} 
w(z)=w_{0}+w_{a}(1-a)=w_{0}+w_{a}z/(1+z)
\label{wz}
\end{equation}
($a$ being the scale factor), although bearing in mind that they are
already a restricted class of dDE models, their state equations being
fully characterized by amplitude and derivative at $z=0$. Through a
Fisher Matrix (FM) approach, we evaluate the errors in WL estimates
for $w_{0}$, $w_{a}$ and $\Omega_m$ (the matter density parameter). We
do so using matter power spectra obtained either from {\sc halofit} or
$N$--body simulations and then we compare the likelihood ellipses on
the parameters obtained with the two approaches. To our knowledge,
this is the first paper in which a Fisher matrix approach is performed
directly on $N$--body outputs rather than on fits or other
approximations.

\begin{figure}
\begin{center}
\includegraphics[scale=0.45]{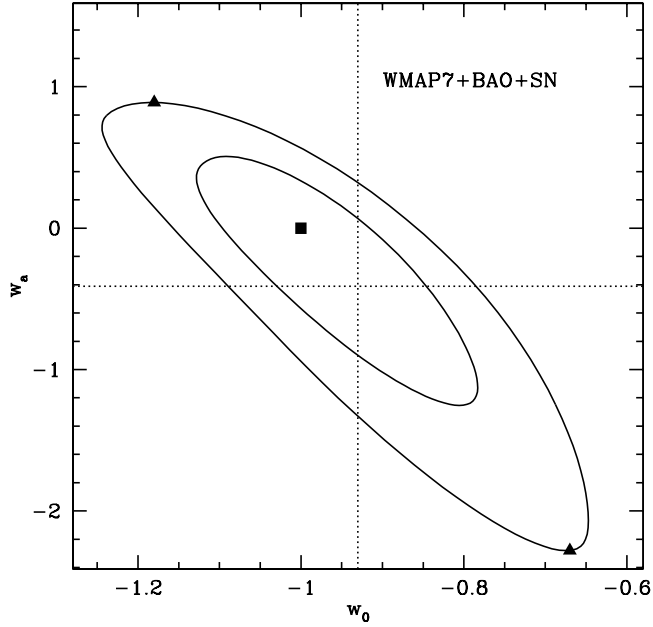}
\end{center}
\caption{Constraints on the time--dependent DE equation of state
(\ref{wz}), derived from WMAP7 spectra combined with SNIa and BAO
updated data \cite{wmap7}. The contours show the 68\% and 95\% CL. The
square point indicate the $\Lambda$CDM model; the triangle points are
the other two reference cosmologies considered in this work: the model
with negative (positive) $w_a$ is also dubbed M1 (M3). The
intersection of the dotted lines is the best fit point in \cite{wmap7}
$w_o=-0.93$, $w_a=-0.41$.}
\label{ellipse}
\end{figure}

We will assume that the shear field is observed in a survey which
approximate the current design of Euclid \cite{euclid2}. This will be
a survey of approximately half sky (20,000 square degrees) with
galaxies at an average depth of $z_m\approx 0.9$ and 40 galaxies per
square arcminute. The redshifts will be evaluated photometrically
assuming a normal distribution with variance $\sigma_z=0.05$ for
errors. The tests are performed assuming as fiducial cosmology
$\Lambda$CDM model ($w \equiv -1$), as well as for two dDE models,
still consistent with the CMB data from WMAP7 \cite{wmap7} and related
data, roughly the most distant models from $\Lambda$CDM that are still
allowed by the data (at 95\% confidence level (CL)~) in the
$w_o$--$w_a$ plane (see figure \ref{ellipse}). In this way we explore
the dependence of our results on the assumed fiducial model.
We can anticipate the conclusion that a straightforward use of {\sc
halofit} is significantly misleading, in quite a few cases.

\section{Weak lensing parameter estimation}
\label{sec:wvar}
The first step to estimate the errors in the parameter measured
through weak lensing techniques, amounts to define the convergence
weak lensing power spectrum, which in the linear regime is identical
to the ellipticity power spectrum \cite{bib6}. This spectrum is a
linear function of the matter power spectrum convolved with the
lensing properties of space. According to \cite{hujain}, if we bin the
galaxies in $n_b$ redshift bins labeled by $i,j=1,...n_b,$ it reads
\begin{equation}
P_{ij}(\ell) = H_{0}^{3}\int_{0}^{\infty} \frac{dz}{E(z)} W_{i}(z)
W_{j}(z)~ P_{nl}  \left( \frac{H_0 \ell}{r(z)},z \right)~.
\label{pijl}
\end{equation}
Here $P_{nl}(k,z)$ is the non-linear matter power spectrum at redshift
$z$ and the $W_i$ are window functions that will be explicitely given
below. Through this paper $n_b=1$, 3 and 5 will be considered, in
order to assess the dependence of the results on the binning procedure
(there is no much gain in going beyond 5 bins \cite{hu}). In the last
two cases, the bin limits $z_i$ are selected so to have the same
number of galaxies per bin. Let us then define the other quantities in
this expression, starting from the redshift dependence of the Hubble
parameter $H(z) \equiv E(z) H_0$ and
\begin{equation}
r(z_1,z_2) = \int_{z_1}^{z_2} \frac{dz}{E(z)}~;
\label{r12}
\end{equation}
accordingly, by referring to the scale factor $a$ and the conformal
time $\tau$, we have that
\begin{equation}
dz/E(z) = H_0(-da/a^2)/(\dot a/a^2) = -H_0 d\tau~,
\end{equation}
so that eq.~(\ref{r12}) also reads
\begin{equation}
r(z_1,z_2) = H_0(\tau_1-\tau_2)~.
\end{equation}
From here we deduce that $r(z) \equiv r(0,z) = H_0[\tau_0 - \tau(z)]$.

Let us then consider also the distribution of the galaxy number per
unit redshift and solid angle. We assume the following simple
parametrized form
\begin{equation}
n(z) = {d^2 N}/{d\Omega\, \, dz} = {\cal C} ~({z}/{z_0})^A 
\exp[-( {z}/{z_0} )^B ]~~~~{\rm
with}~~~~ {\cal C} = (B/z_0)~ \Gamma^{-1}[(A+1)/B]
\label{nz}
\end{equation}
where $\Gamma(x)$ is the Gamma function. Here we choose $A=2$,
$B=1.5$, so that ${\cal C} = 1.5/z_0$. If $z_m=0.9$ is the average
redshift, one then has $z_0 = z_m/1.412$. This choice of values is a
good approximation to the observed selection function of a survey like
Euclid \cite{euclid2}. By integrating the distribution $n(z)$ across
the depth of the $j$--th bin $\Delta z_j$ we have the angular density
in steradians
\begin{equation}
 n_j =3600 (180/\pi)^2 \int_{\Delta z_j} dz~n(z)~,
\label{nj}
\end{equation}
an expression to be used in the next section. 

The distribution (\ref{nz}) is then considered within the limits of
the redshift bins. In order to take into account the discrepancies
between the photometric redshift in use, and the actual galaxy
redshift, we filter the distributions using window functions
\cite{amref,amref2}
$$\Pi_i (z) = \int_{z_{ph,i}}^{z_{ph,i+1}} dz' ~ \frac{1}{\sqrt{2 \pi}~
\sigma(z)} \exp \left(-\frac{(z - z')^2}{2 \sigma^2 (z)} \right) =
~~~~~~~~~~~~$$
\begin{equation}
~~~~~~~~ = \frac{1}{2}
[Erf(z_{ph,i+1}-z)/\sqrt{2}\sigma(z)-Erf(z_{ph,i}-z)/\sqrt{2}\sigma(z)]
\end{equation}
with $\sigma(z) = \sigma_z~(1+z)$, $\sigma_z=0.05$ (see, e.g.,
\cite{amara,sigmaz}, for the motivation of this parameter choice).

\begin{figure}
\begin{center}
\includegraphics[scale=0.45]{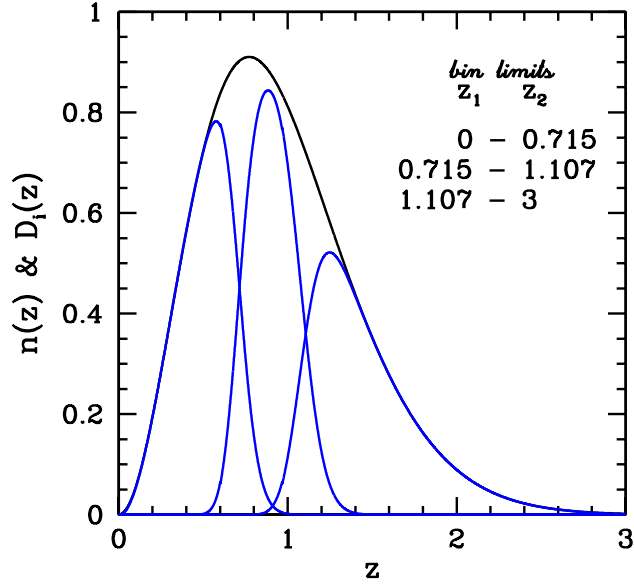}
\end{center}
\caption{$D_i(z)$ distributions in the 3--bin case. The bin limits
$z_1$ \& $z_2$, shown in the frame, are selected to have the same
number of galaxies in each bin. In the 5--bin case the bin limits
$z_i$ are 0.560, 0.789, 1.019, 1.324, 3.00~. The background black line
is $n(z).$ }
\label{windows}
\end{figure}

In figure \ref{windows} we show the radial galaxy distribution and the
distributions
\begin{equation}
D_i(z) = n(z) \Pi_i(z)
\label{diz}
\end{equation}
in the case of three bins. They must then be
\begin{figure}
\begin{center}
\includegraphics[scale=0.45]{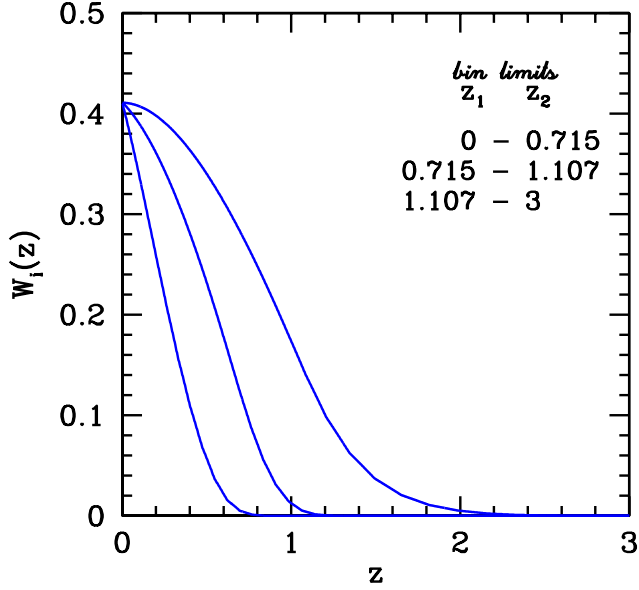}
\end{center}
\caption{$W_i(z)$ window functions adopted for the 3--bin case.}
\label{wiz}
\end{figure}
normalized, yielding the distributions
\begin{equation}
\delta_{i}(z) = D_{i}(z) \bigg/ \int_{0}^{\infty}D_{i}(z')dz'
\end{equation}
wherefrom we derive the functions
\begin{equation}
\label{WI}
W_i (z) = \frac{3}{2} \Omega_m F_i(z) (1+z)
~~~~~~{\rm with} ~~~~~~
F_i (z) = \int_{\Delta z_i} dz'~\delta_i(z') r(z,z')/r(z')~,
\end{equation}
shown in figure \ref{wiz}.

From the power spectrum $P_{ij}(\ell)$, we obtain the covariance
matrix
\begin{equation}
C_{jk} = P_{jk} + \delta_{jk} \left\langle \gamma_{int}^{2}
\right\rangle n_{j}^{-1}~,
\end{equation} 
including the effect of the r.m.s. intrinsic shear $\gamma_{int}$, for
which we assume $\left\langle \gamma_{int}^{2}\right\rangle ^{1/2} =
0.22$ \cite{amara}; $n_j$ is given by eq.~(\ref{nj}).

The above expressions enable us to make use of the Fisher matrix
formalism, which provides lower limits to the errors on the
cosmological parameters one aims to measure. Its basic tool is the
likelihood function, yielding the probability that a model, fixed by a
set of parameters $p_\alpha$, gives the set of data {\bf x}.

The Fisher matrix for tomographic weak lensing reads \cite{hujain}
\begin{equation}
F_{\alpha\beta} = f_{sky}\sum_{\ell}^{\ell_{max}} \frac{(2\ell+1)
\Delta\ell}{2} \frac{\partial C_{ij}}{\partial p_{\alpha}} C_{jk}^{-1}
\frac{\partial C_{km}}{\partial p_{\beta}} C_{mi}^{-1}~.
\label{derfish}
\end{equation}
The basic ingredients of this expressions are the $C_{ij}$ components;
$f_{sky}$ is the sky fraction covered by the experiment (in our case
$f_{sky}= 0.5$); the cosmological parameters $p_{\alpha}$ we shall
consider are $w_o$, $w_a$ and $\Omega_m$, so that $\partial/\partial
p_{\alpha}$ are partial derivatives with respect to them.

All the other parameters are fixed, in order to keep the number of
$N$--body simulations to a manageable level, by assuming data to
provide us the other parameters with negligible errors. These
parameters include $\Omega_b$, $n_s$, $h$ (baryon density parameter,
primeval spectral index of scalar fluctuations, adimensional Hubble
parameter), plus a parameter setting the spectral normalization.

If data yield non--linearly evolved amplitudes, it is from them that
we should guess suitable initial conditions at $z_{in}=24$, when
starting the simulations, in the linear regime. Here we shall need to
improve on the standard normalization procedure which neglects the
mild discrepancy between the r.m.s. matter density fluctuation on the
comoving scale $R_8 = 8\, \, h^{-1}$Mpc, taken from data, and the
linear expression
\begin{equation}
\sigma^2_{8,l} (z) = (2\pi^2)^{-1} \int_0^\infty dk\, \, 
A\, k^{n_s + 2} {\cal T}^2 (k,z) W^2(kR_8)~;
\label{s8l}
\end{equation}
here $A$ is the primeval fluctuation amplitude at a suitable very high
$z_0$ value, ${\cal T}(k,z)$ is the linear transfer function from such
$z_0$ value to $z$, $W(x)$ is a window function, for which we can use
the Fourier transform of a top--hat filter: $W(x) = (3/x^3)(\sin
x-x\cos x)\, \, \, $. This mild discrepancy, infact, is $\cal O$$(10\,
\%)$, too much for the signal we are chasing.

Normalizing models to have the same $\sigma_{8,nl}$, as we then need
to do, is clearly harder and includes technical aspects which will be
further discussed below.

Should we forget this point, and normalize all models so to have the
same $\sigma_{8,l}$ at $z=0$, the derivatives with respect to $w_o$,
$w_a$ or $\Omega_m$ are largely dominated by the normalization shift
at $z=0$, as the $\sigma_{8,l}$--$\sigma_{8,nl}$ shift itself depends
on $w_o$, $w_a$ and $\Omega_m$. This would confuse the $z$ dependence
of the growth factor, through the observational $z$--range, i.e. the
main observable that lensing measures want to exploit.

Once the fiducial models are chosen, the derivatives in
eq.~(\ref{derfish}) are evaluated by extracting the power spectra from
the simulations of models close to the fiducial ones, obtained by
considering parameter increments $\pm 5\, \%$.

Our task will then amount to find the regions above a given confidence
level. They are $M$--dimensional ellipsoids, about the $M$ parameter
values corresponding to fiducial models, that we shall then project
onto 2--dimensional subspaces spanned by parameter pairs, so
determining constant $\chi^2$ contours. In the Fisher Matrix that we
adopt, such {\it marginalized likelihood} contours will be ellipses
centered on the fiducial cosmology.

In a generic case, when the $M$ parameters are dubbed $X_\alpha$
($\alpha=1,2,..,M$), the ellipse equation in the $X_\mu,X_\nu$ plane
($n=2$ subspace) reads
\begin{equation}
\left( X_\mu X_\nu \right) \left[
\begin{matrix}
F^{-1}_{\mu \mu} & F^{-1}_{\mu \nu} \\
F^{-1}_{\nu \mu} & F^{-1}_{\nu \nu} 
\end{matrix}
\right]^{-1} \left( 
\begin{matrix}
X_\mu \\ X_\nu
\end{matrix}
\right) = 
\Delta \chi^2 (CL,n) ~.
\end{equation}
Here $\Delta \chi^2$ sets the confidence level, being
\begin{equation}
\Delta \chi^2 (68.3\, \%,~n_p) = 2.30~,~~~~
\Delta \chi^2 (95.4\, \%,~n_p) = 4.61~,~~~~
\label{CL}
\end{equation}
for $n_p=2~.$

\section{Cosmological models}
In figure \ref{ellipse} the settings of the fiducial models are
shown. The corresponding values of $w_o$ and $w_a$, are reported in
Table \ref{tab1}. We also take $\Omega_{m}=0.274$, $h=0.7$,
$~\sigma_{8,l}=0.81 $, $n=0.96$ (matter density parameter, Hubble
parameter in units of 100 km/s/Mpc, {\it linear} r.m.s matter
fluctuation amplitude at $z=0$ on the scale of 8~$h^{-1}$Mpc, primeval
spectral index, respectively) consistently with WMAP7 outputs. We
shall discuss below why (slightly) different $\sigma_{8,l}$ values
need also be considered.

About the fiducial models we select models characterized by $\pm 5\,
\%$ increments of $w_o$ and $w_a$, to perform numerical
derivatives. In Table \ref{tab1} the list of these models is reported.
\begin{table}[t!]
\centering
\begin{tabular}{c c c c c c}
\hline
\multicolumn{2}{c}{$\Lambda$CDM} & \multicolumn{2}{c}{M1} & \multicolumn{2}{c}{M3} \\
\hline
\hline
$w_0$ & $w_a$ & $w_0$   & $w_a$ & $w_0$ & $w_a$  \\
 -1   & 0     & -0.67   & -2.28 & -1.18 & 0.89 \\
-1.05 & 0     & -0.7035 & -2.28 & -1.239& 0.89 \\
-0.95 & 0     & -0.6365 & -2.28 & -1.121& 0.89 \\
 -1   & +0.05 & -0.67   & -2.394& -1.18 & 0.9345 \\
 -1   &-0.05  & -0.67   & -2.166& -1.18 & 0.8455 \\
\hline
\end{tabular}
\caption{List of values of $w_o$ and $w_a$ used to perform numerical
derivatives.}
\label{tab1}
\end{table}

When $\Lambda$CDM is the fiducial model, we consider also derivatives
with respect to the $\Omega_m$ parameter. In this case all the
procedure deals with $\Lambda$CDM cosmologies; we shall then work out
the spectra also for $\Lambda$CDM models with $\Omega_{m}=0.2603$ and
0.2877.

For $\Lambda$CDM and all nearby models, two different initial seeds
were also considered, to test the dependence on initial conditions,
finding that Fisher Matrix results are almost insensitive to
it. Accordingly, for the other fiducial models, only one seed is
used. Altogether we run 24 model simulations.

\section{Simulations and their analysis}
\begin{figure}
\begin{center}
\includegraphics[scale=0.57]{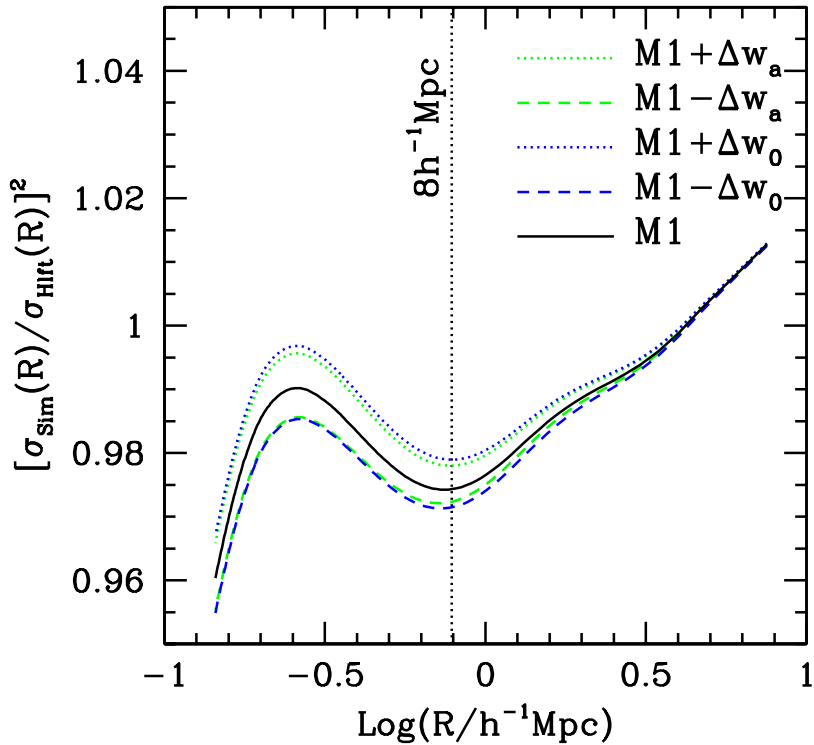}
\end{center}
\caption{$R$--dependence of the ratio between the $\sigma^2(R)$'s at
$z=0$ obtained from simulations, for models in the neighborhood of M1;
they are normalized to the respective mass variances obtained using
{\sc halofit}, however almost identical for all these models (see
text).}
\label{sigrat}
\end{figure}
\begin{figure}
\begin{center}
\includegraphics[scale=0.65]{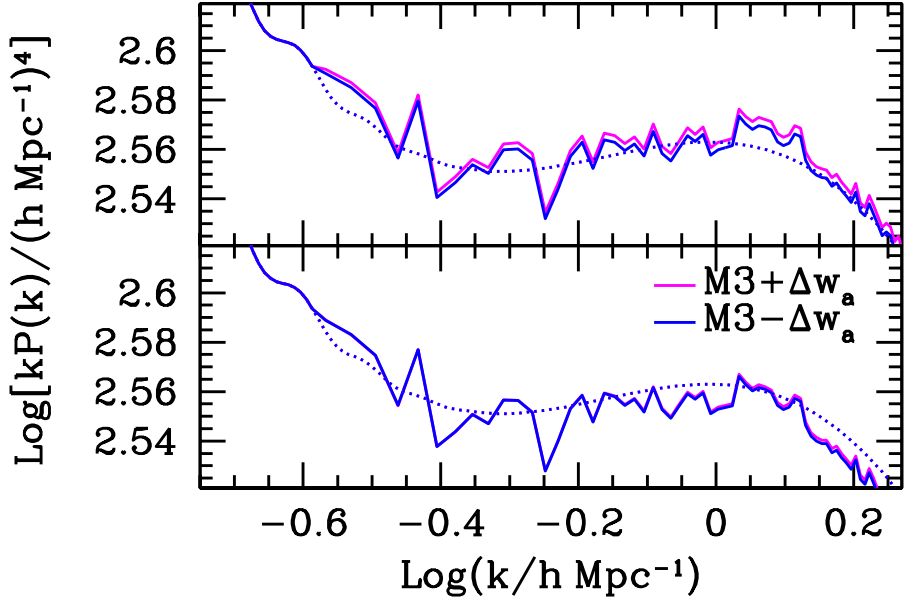}
\end{center}
\caption{Connection between {\sc halofit} and simulation spectra at
the low--$k$ end. The upper (lower) panel shows the resulting spectra
when normalizing to $\sigma_{8,l} (z=0)$ ($\sigma_{8,nl} (z=0)$). The
dotted line is {\sc halofit}.}
\label{baco}
\end{figure}
$N$--body simulations are performed by using a modified version of
PKDGRAV \cite{pkdgrav} able to handle any DE state equation $w(a)$. To
this aim, $N^3 = 256^{3}$ particles, representing CDM and baryons, are
set in a box with side $L = 256\, h^{-1}$Mpc. Particle masses are
therefore $m_{c} = \rho_{o,cr} \Omega_m (L/N)^3 = 7.61 \times 10^{10}
h^{-1} M_{\odot}$ ($\rho_{o,cr}:$ critical density). Softening is 1/40
of the initial intra--particle distance, yielding $\epsilon \simeq
25~h^{-1}$kpc (wavenumber $\kappa = 2\pi/\epsilon \simeq 150\,
h\,$Mpc$^{-1}$).

Transfer functions generated using the CAMB package are employed to
create initial conditions, with a modified version of the PM software
by Klypin and Holzmann \cite{klypinholzmann}, also able to handle
suitable parameterizations of DE. Simulations are started at
$z_{in}=24$. The initial density field is obtained from the same
random numbers for all the simulations. For the $\Lambda$CDM fiducial
we generate each simulation with two sets of initial random numbers.

Matter power spectra are obtained by performing a FFT (Fast Fourier
Transform) of the matter density fields, that we computed from the
particles distribution, through a Cloud--in--Cell algorithm, by using
a regular grid with $N_{g}=2048$. This allows us to obtain non--linear
spectra in a large $k$--interval.

In particular, our resolution allows to work out spectra up to $k
\simeq 10\, h$Mpc$^{-1}$. However, for $k > 2$--$3\, h\, $Mpc$^{-1}$
neglecting baryon physics is no longer accurate \cite{3p}. Although
both simulations and {\sc halofit} refer to purely gravitational
dynamics, we wish to examine only physically significant scales.
Therefore, we consider WL spectra for $\ell < 2000$ only, so limiting
the contribution of scales $k > 2\, h\, $Mpc$^{-1}$ below $\simeq 5\,
\% \, ,$ at most.

Let us now discuss the normalization problem. In Figure
\ref{sigrat} we show the dependence on scale of the r.m.s. of matter
density fluctuations (mass variance), as obtainable through the
expression
\begin{equation}
\sigma^2 (R) = \sum_{k} \Delta k_k\, P(k)\, k^2\, W^2(kR)
\label{s3r}
\end{equation}
where $P(k)$ is the full non-linear spectrum. When applied to
simulations, we denote the variance as $\sigma^2_{sim}(R) $. Plotting
the scale dependence for $\sigma^2_{sim} (R)$, rather than for the
simulation spectra $P_{sim}(k)$, smears out the numerical oscillations
that spectra unavoidably exhibit.

Figure \ref{sigrat} allows us to focus on the discrepances between the
model M1 and the nearby models used to perform derivatives about it.
The spectra shown in the figure were obtained by normalizing initial
conditions so that their $\sigma_{8,l}$ at $ z=0$ coincide.  As a
consequence the whole linear spectra of M1 and nearby models (not
shown in the Figure) are nearly coincident, at $z=0$, since the models
share all parameters except the DE state equation, which has a very
weak impact on the linear spectrum shape.  If one uses {\sc halofit}
to correct the linear spectra, the $z=0$ non-linear spectra would
still coincide, since {\sc halofit} only depends on $\sigma_{8,l}$ and
on $\Omega_{de}$, not on $w$.  In contrast, Figure \ref{sigrat}
confirms that the simulation spectra differ. Accordingly, taking the
Fisher matrix derivatives fixing $\sigma_{8,l}$ is different from
fixing $\sigma_{8,nl}$. Besides of the points made above, here we
choose to normalize to the same non-linear variance $\sigma_{8,nl}$
also because this is the primary observational quantity and we expect
that future data will put direct constraints on it, resulting in a
strong prior in the Fisher matrix.

The price to pay is that producing simulations with different
cosmological parameters but coincident $\sigma_{8,nl}$ is not a
trivial task and requires a trial-and-error search of initial
conditions for the $N$-body simulations.

There is then another question, concerning the low--$k$ spectrum. The
point is that the integration (\ref{pijl}) unavoidably extends down to
$k$'s which are still close to (or fully in) the linear regime. In
principle, dealing with such scales should not create
difficulties. However, the spectra obtained from our simulations, run
in a box of 256$\, h^{-1}$Mpc, become increasingly noisy and discrete
when the linear regime is approached. To solve this problem we
extrapolate our numerical spectra with the linear theoretical ones at
scales larger than a certain threshold.

The problem is then where and how to connect numerical spectra with
theoretical ones. Let then be $k_p = 2p\pi/(256\, h^{-1}$Mpc), being
$p$ an integer; we use simulation spectra when $\Delta k_p/ k_p $
shifts below 0.1~, i.e. for $\log(k/h\, {\rm Mpc}^{-1}) > -0.61~.$ At
lower $k$'s {\sc halofit} spectra are used.

Figure \ref{baco} (upper frame) is the result of this operation, when
done using $M3 \pm \Delta w_a$ spectra characterized by the same
$\sigma_{8,l} (z=0)$. Spectral shifts of this kind affect the ratios
between derivatives with respect to cosmological parameters (in this
case, $w_a$). In the same Figure (lower frame) we show the result of
the connection after renormalizing spectra to the same
$\sigma_{8,nl}$. Some scale dependent discrepancy still remains, 
but the main signal comes then from the different redshift dependence
of model spectra, rather than from the shift among $\sigma_{8,nl}$.

\begin{figure}
\begin{center}
\includegraphics[scale=0.65]{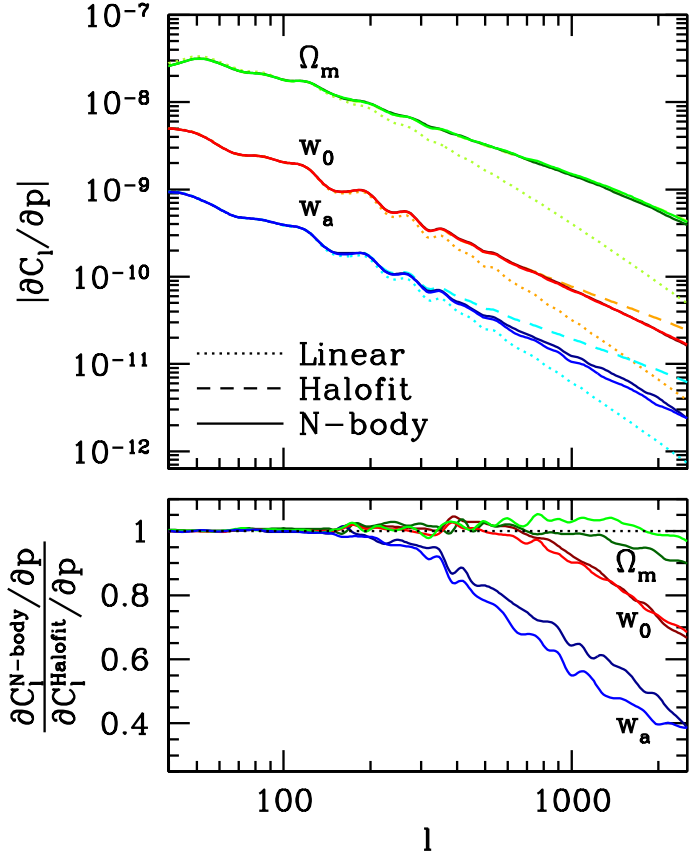}
\end{center}
\caption{Derivatives of $C_\ell$, in the case of a single bin, for $20
< \ell < 2000$, with respect to the parameters $\Omega_m$, $w_0$ and
$w_a$, when the fiducial model is $\Lambda$CDM. In the upper frame we
show their different behavior, for $N$--body simulations, {\sc
halofit} and linear spectra. In the lower frame we show the ratio
between derivatives obtained from $N$--body simulations and {\sc
halofit}. For these models we run simulations with two different
seeds. Derivatives obtained from each of them are in the same colors,
using a darker tonality for the seeds considered just for these
models.}
\label{derivative}
\end{figure}
\begin{figure}
\begin{center}
\includegraphics[scale=0.5]{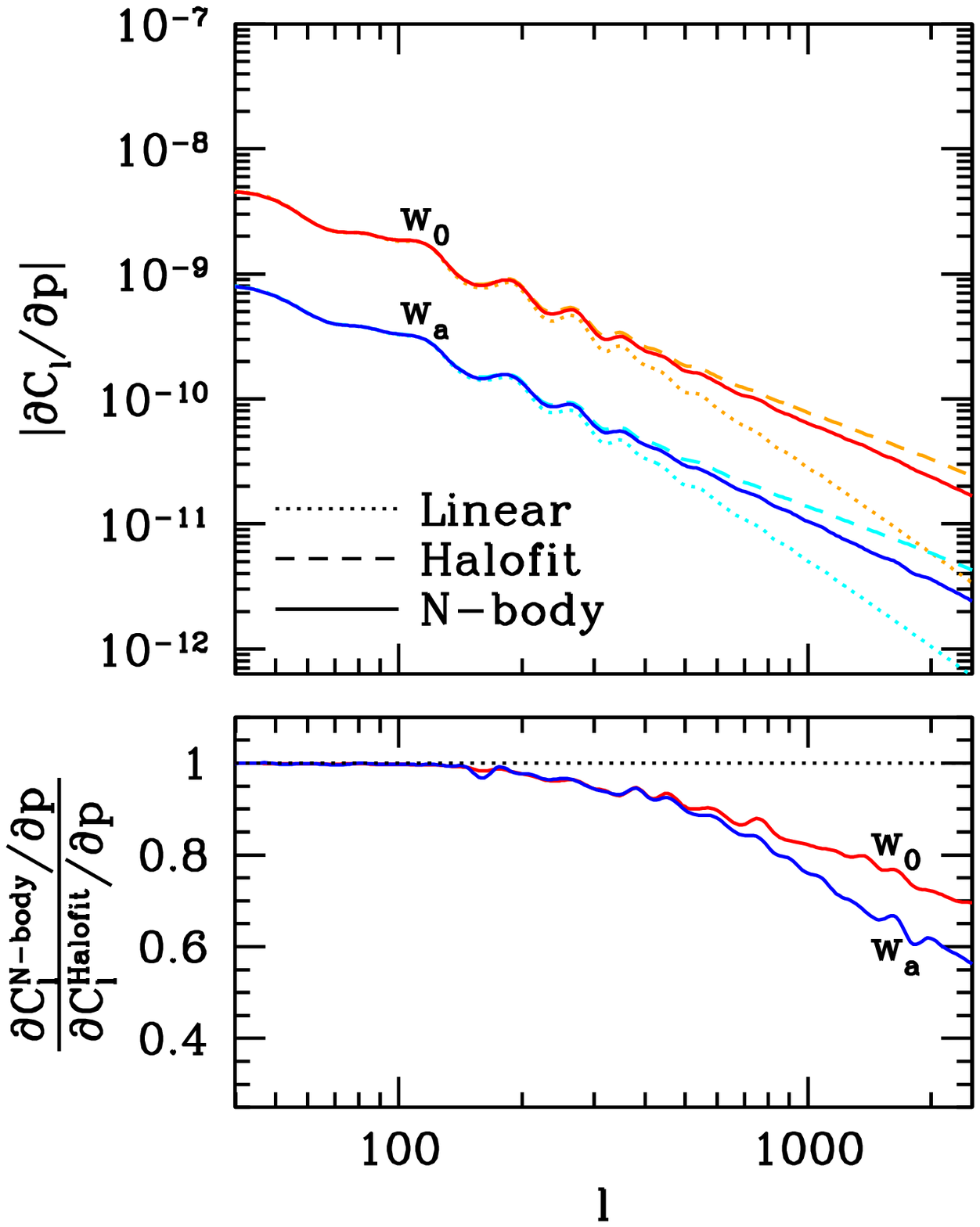}
\includegraphics[scale=0.5]{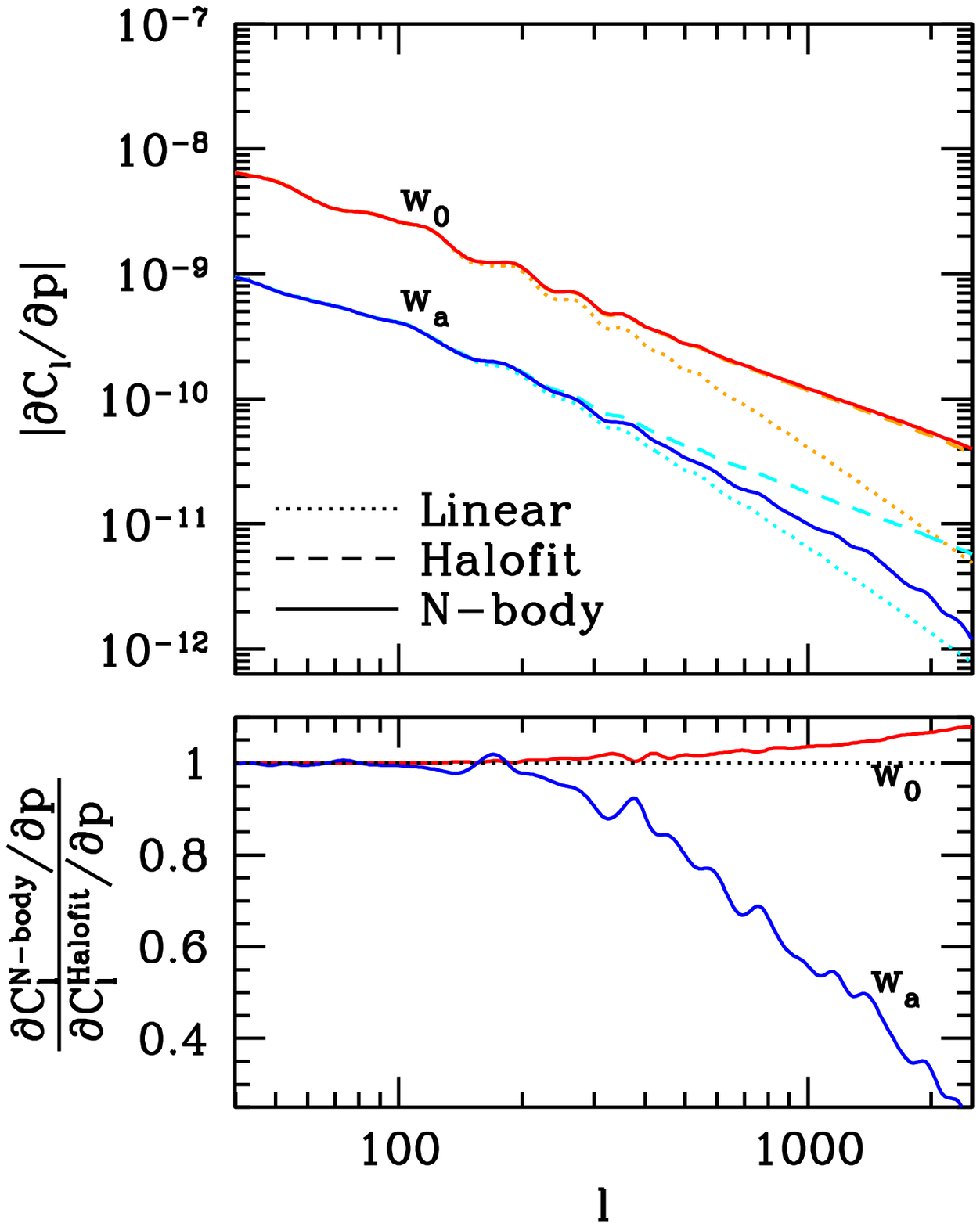}
\end{center}
\caption{Derivatives of $C_\ell$, as in the previous figure, for M1
and M3 fiducial models (l.h.s. \& r.h.s., respectively), omitting the
parameter $\Omega_m$, which is not considered in these cases. }
\label{derivative12}
\end{figure}

\section{Predictions on parameter errors}

\subsection{Derivatives}
The next step is evaluating the derivatives in eq.~(\ref{derfish}).
For the sake of illustration we plot the results of such
differentiation in the case of a single bin, by considering: (i) the
spectra obtained from simulations, as illustrated in previous section;
(ii) {\sc halofit} spectra; (iii) purely linear spectra.

In figure \ref{derivative} we show the derivatives and, in the lower
frame, compare the results of the cases (i) and (ii), when the
fiducial model is $\Lambda$CDM. The results obtained starting
simulations from two different seeds are also compared. In figure
\ref{derivative12} we similarly show derivatives when the fiducial
model is M1 or M3. One seed only is considered there, and we consider
the parameters $w_0$ and $w_a$ only.

The effects of BAO's as well of irregularities in simulation spectra
are clearly visible in all plots. There is however a clear difference
between the derivative with respect to $\Omega_m$, in figure
\ref{derivative}, and any other one. For both seeds the ratio between
its value for simulations and {\sc halofit} is close to unity, with a
discrepancy exceeding $\sim 5\, \%$ just for $\ell > 1500$, when
contributions from $k$ values above the range of {\sc halofit}
validity become non--irrelevant. This is a further confirmation of the
coherence between $N$--body simulations and {\sc halofit}, for
$\Lambda$CDM models, the only ones involved in the differentiation.

Still in figure \ref{derivative}, we notice that differences between
seeds mostly stays within 6--7$\, \%$, even in the case of the
parameter $w_a$, which is the most sensitive to noise effects.

The case M1, shown in figure \ref{derivative12}, exhibits then a
peculiarity, with respect to all other cases. Here, the ratios of
derivatives with respect to $w_0$ and $w_a$ almost overlap up to $\ell
\simeq 500,$ while their trend remains similar even up to $\ell =
1000~.$ The effect is peculiar of M1, and recurs also when more bins
are considered. For $\Lambda$CDM, the behaviors of the two ratios are
significantly discrepant. In the case M3, finally, the two ratios
exhibit neatly different behaviors. This reflects onto the shape of
the likelihood ellipses which are one of the main results of our work.

\subsection{Likelihood ellipses: \texorpdfstring{$\Lambda$}{L}CDM 
neighborhood}

\begin{figure}[t!]
\begin{center}
\includegraphics[scale=0.4]{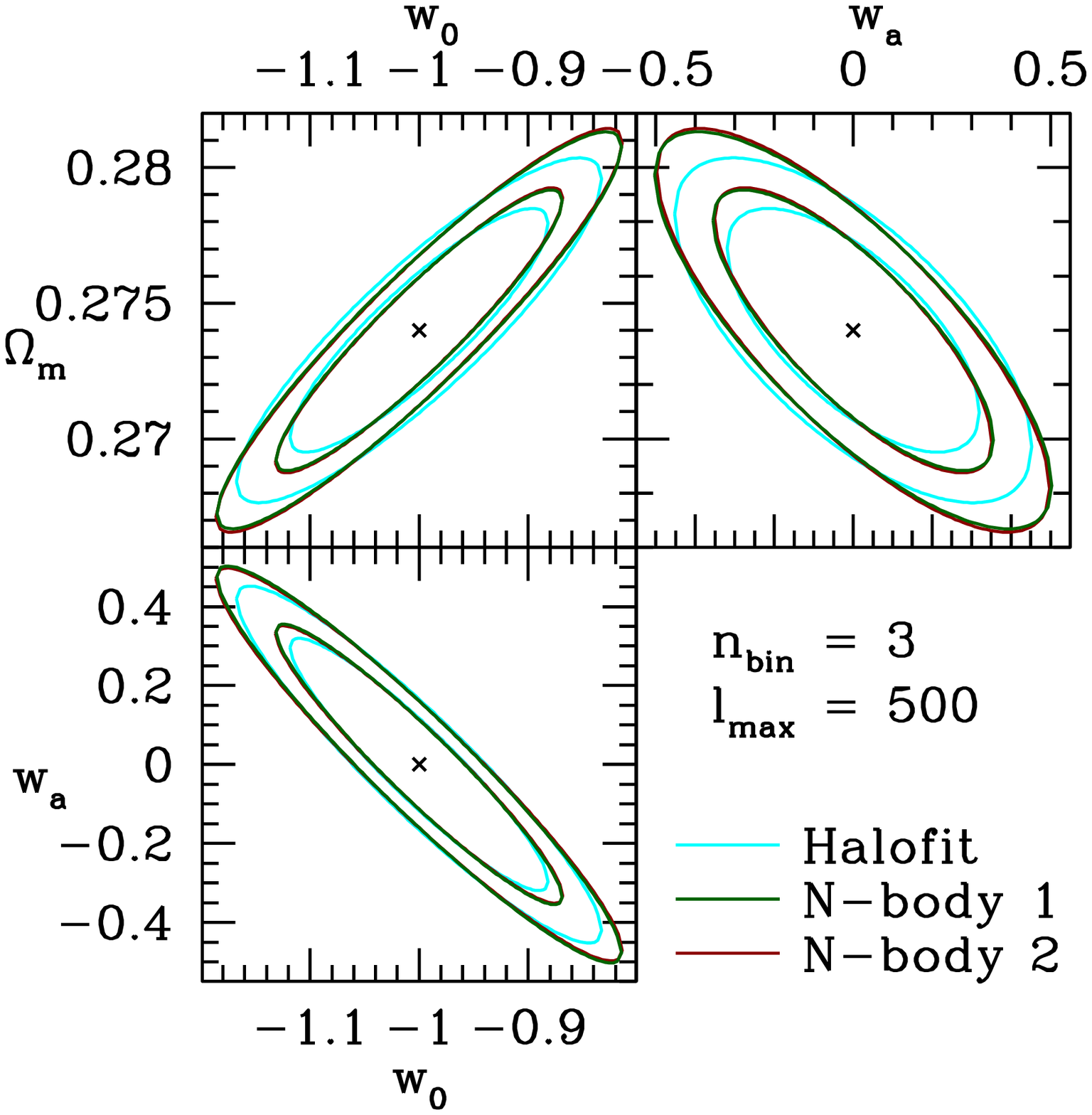}
\hskip .5truecm
\includegraphics[scale=0.4,height=6.95truecm,width=6.88truecm]{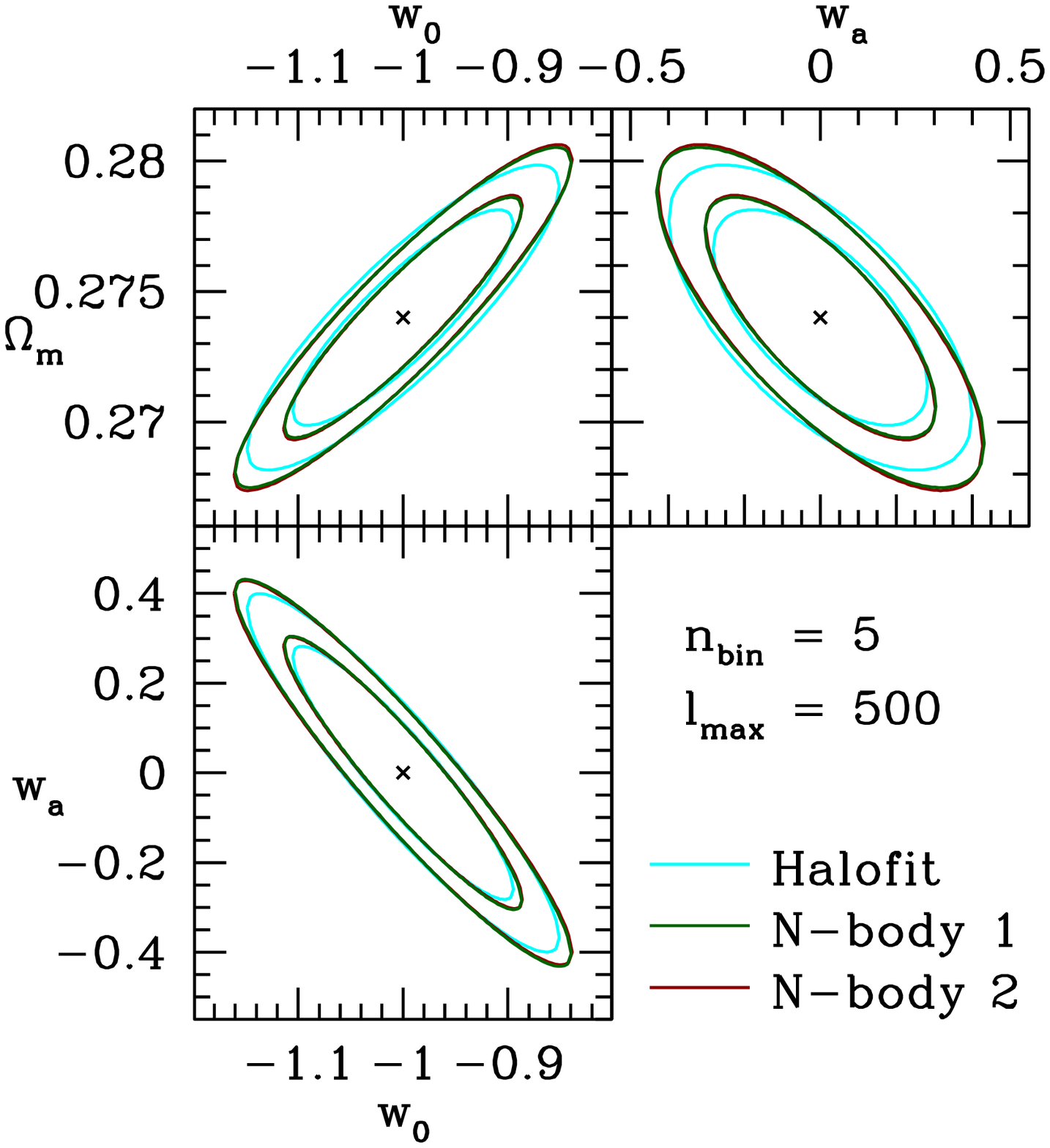}
\vskip .4truecm
\includegraphics[scale=0.4]{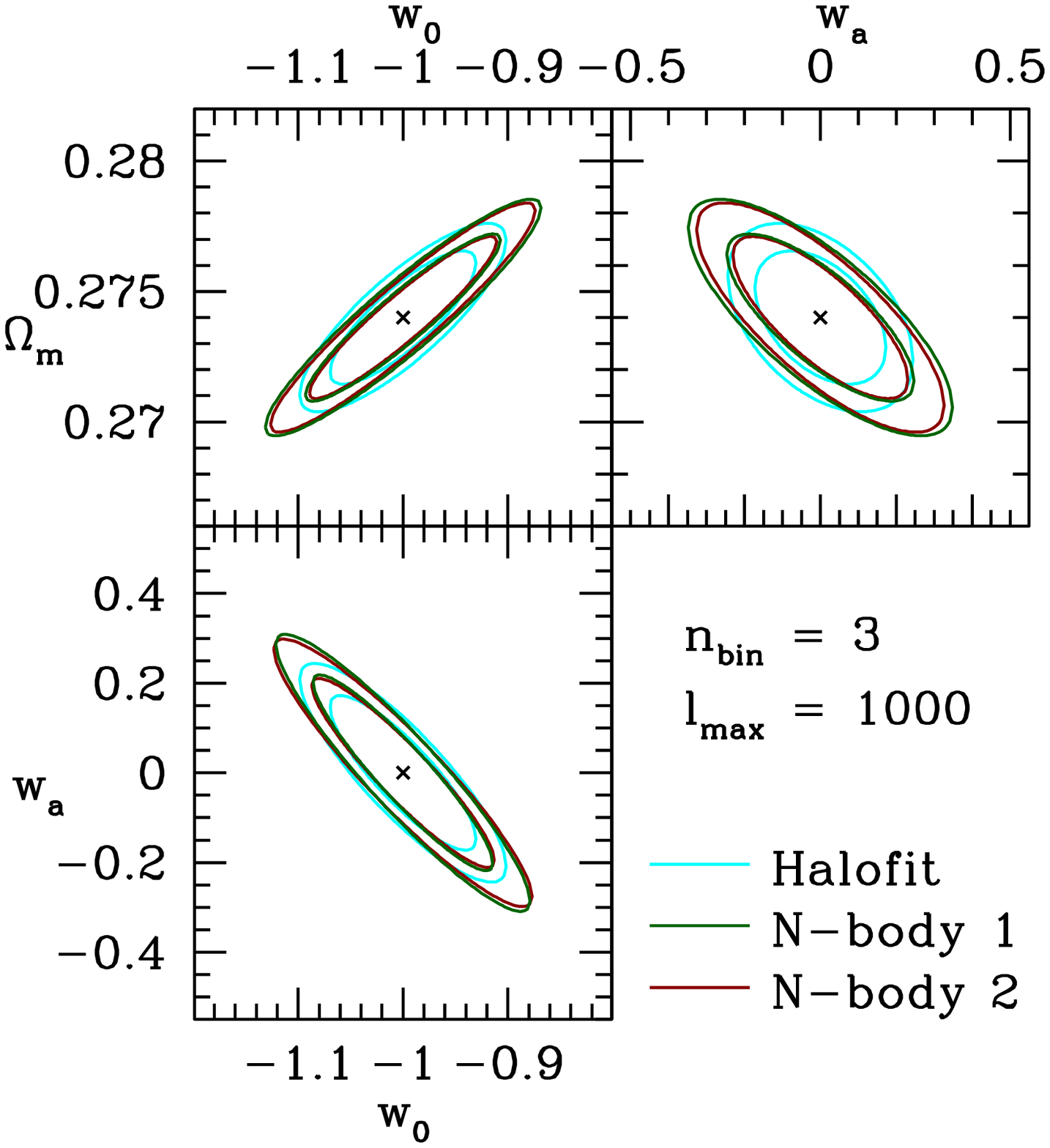}
\hskip .5truecm
\includegraphics[scale=0.4]{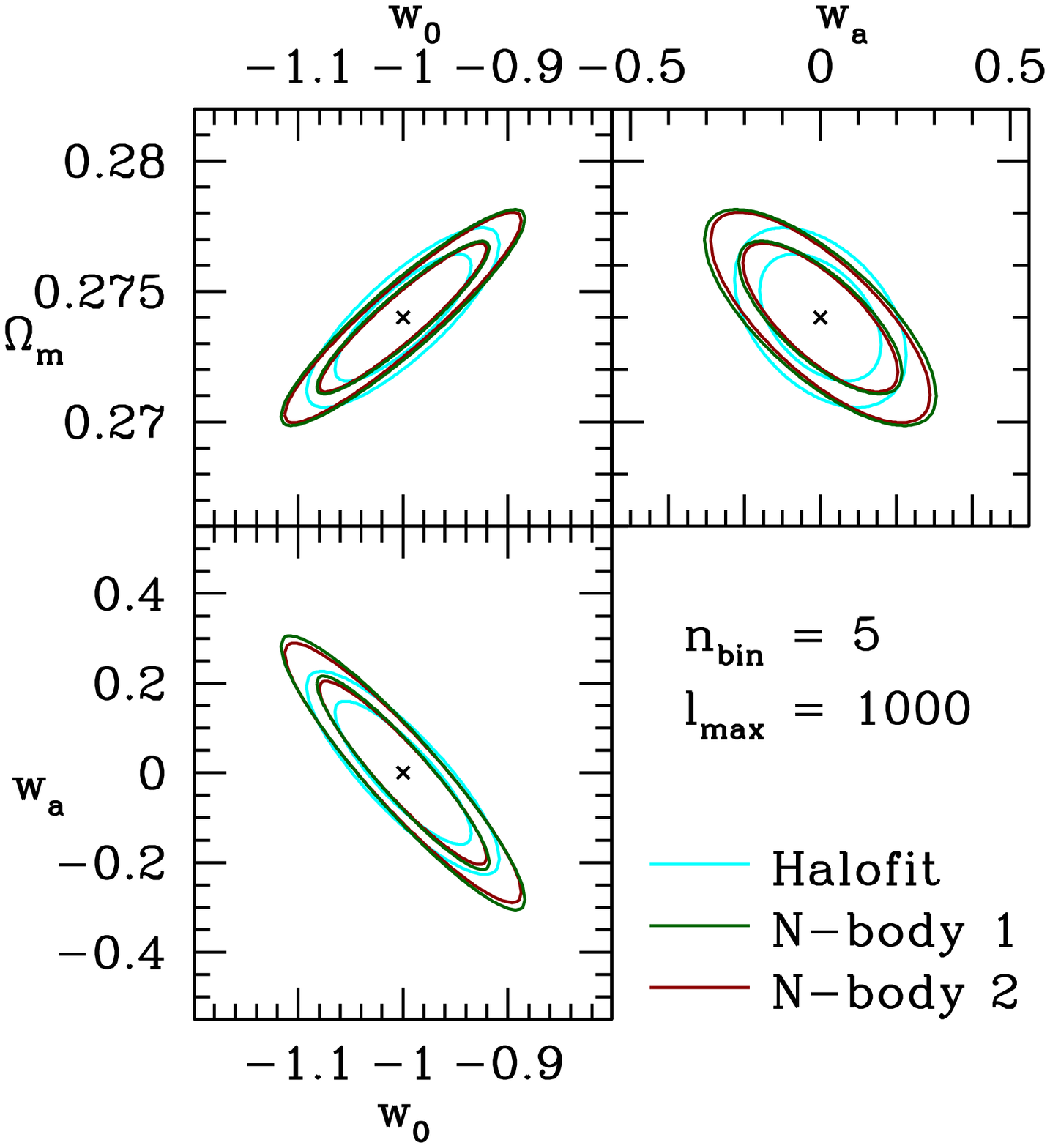}
\end{center}
\caption{Likelihood ellipses, showing 65$\, \%$ and 95$\, \%$ CL
contours, for all parameter pairs, including signals up to either
$\ell \simeq 500$ or 1000 for the $\Lambda$CDM fiducial. In the former
case no significant discrepancy between {\sc halofit} and simulation
results can be appreciated. Discrepancies become more relevant in the
latter case, both for 3 and 5 bins. Notice also that the two
simulation seeds yield almost overlapping outputs.}
\label{lcdm}
\end{figure}

\begin{figure}[t!]
\begin{center}
\includegraphics[scale=0.5]{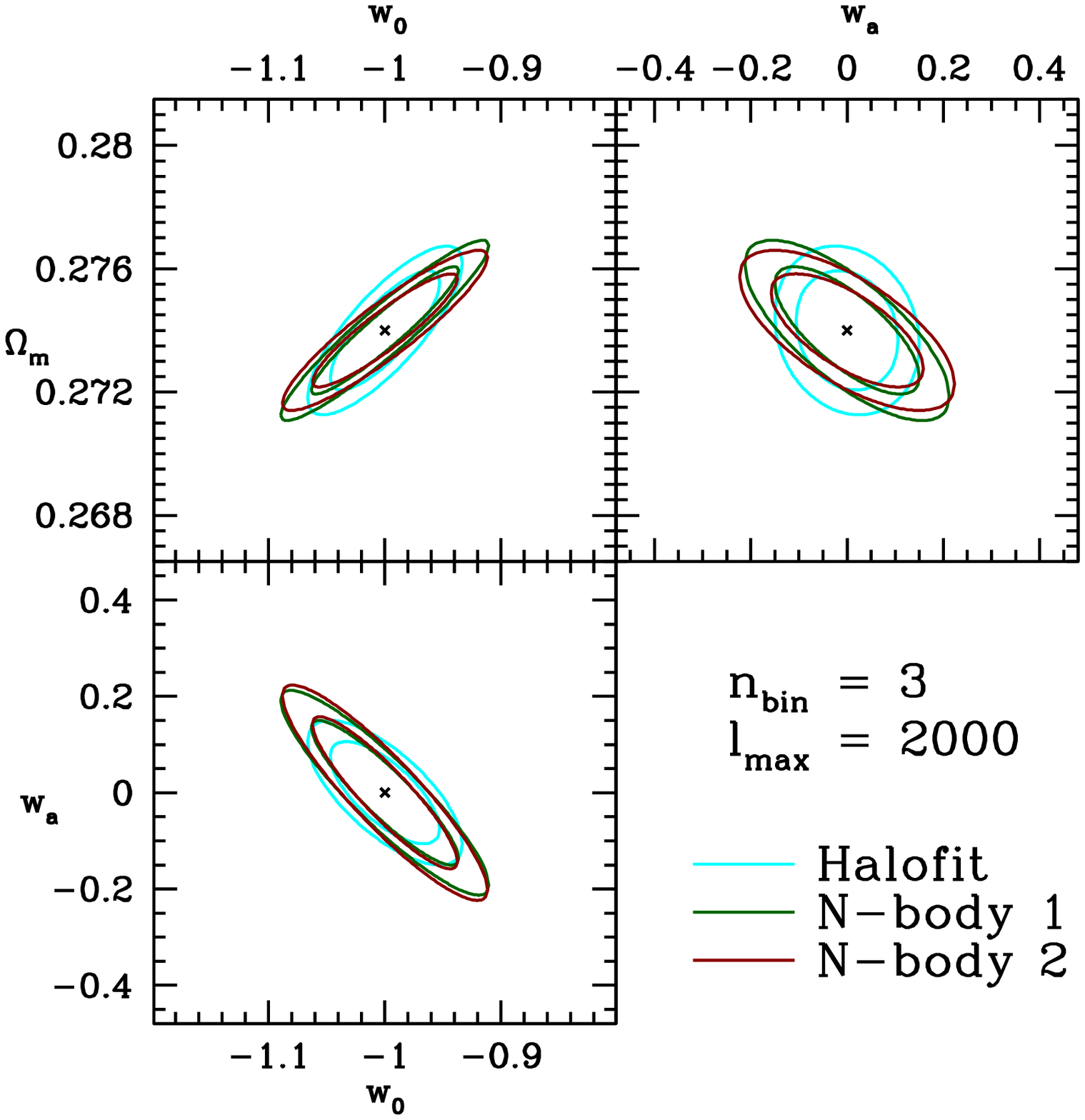}
\includegraphics[scale=0.5]{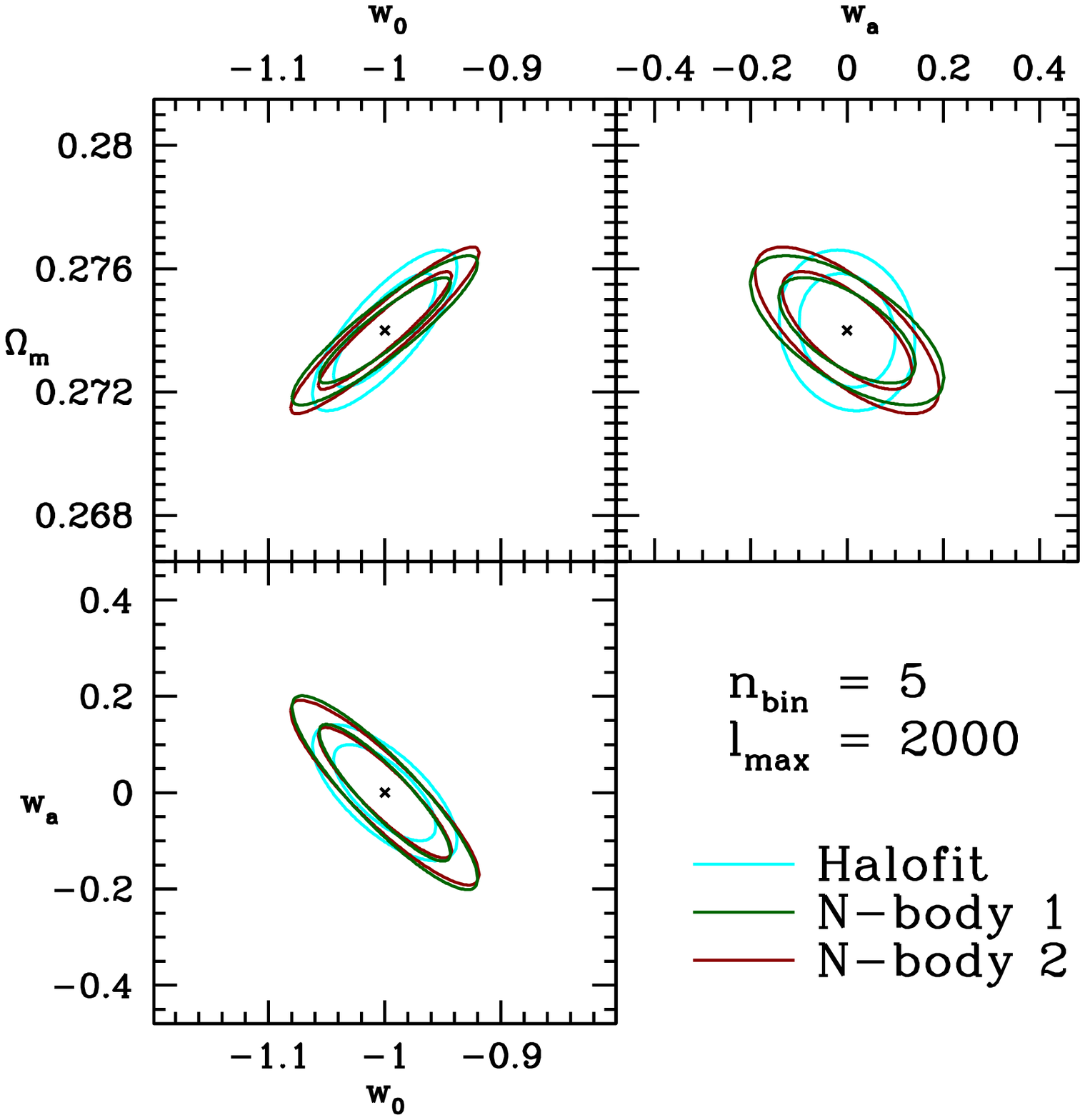}
\end{center}
\caption{Likelihood ellipses, showing 65$\, \%$ and 95$\, \%$ CL
contours, for all parameter pairs, considered including signals up to
$\ell \simeq 2000$ for the $\Lambda$CDM fiducial. Here simulations and
{\sc halofit} yield significantly different outputs.}
\label{lcdm2000}
\end{figure}

In figures \ref{lcdm} and \ref{lcdm2000} we show the confidence
ellipses, when the fiducial model is $\Lambda$CDM, in the cases of 3
or 5 bins and using $C_\ell$ up to $\ell = 500$, 1000 or 2000.

Up to $\ell \simeq 500$, i.e. including just a mildly non--linear
scale range, discrepancies between {\sc halofit} and simulations
already indicate an underestimate of errors in the {\sc halofit}
case. Discrepancies approach a factor 2 when we go up to $\ell =
1000$, both for 3 and 5 bins. They become quite significant in the
$\ell = 2000$ case. In the latter cases we observe a modification of
the likelihood contours, indicating a different correlation between
the errors on different parameters. This includes the
$w_a$--$\Omega_m$ correlation and means that a neglect of the degrees
of freedom of DE state equation leads to errors on the estimate of the
density parameter $\Omega_m$.

Meanwhile, the discrepancy between simulations started from different
seeds are much smaller. Different seeds yield discrepancies still
visible in derivatives (see figure \ref{derivative}), but the
integrations needed to pass from them to likelihood ellipses smear
them out. In fact, the shaping of the ellipses is dominated by the
$z$--dependence of the spectral behavior, and here is where {\sc
halofit} is apparently unable to meet simulation results. The changes
in the confidence ellipses are also quantified in Table \ref{tab2}.

Incidentally, let us notice that the improvements, when passing from
the 3-- to the 5--bin case, are more significant when a greater $\ell$
is considered.

\begin{table}[t!]
\centering
\begin{tabular}{c c c c}
\hline
&&{\sc halofit}&sim. average\\
\hline
\hline
$w_0$--$w_a$  & $\theta$                   & $-71.4^o$  & $-70.1^o$ \\
 plane        & axial~ratio~(95\%\, CL)    & 3.27       & 6.22      \\
              & 1/area(95\%\, CL)          & 50.2       & 43.4      \\
              & correlation                & -0.69      & -0.88     \\
\hline
$w_0$--$\Omega_m$  & $\theta$                   & $1.90^o$  & $1.71^o$ \\
 plane        & axial~ratio~(95\%\, CL)    & 39.3       & 75.7      \\
              & 1/area(95\%\, CL)          & 3.22 $\times 10^3$  & 
3.92 $\times 10^3$  \\
              & correlation                & 0.79      & 0.92     \\
\hline
$\Omega_m$--$w_a$  & $\theta$                   & $-0.15^o$  & $-0.49^o$ \\
 plane        & axial~ratio~(95\%\, CL)    & 54.7       & 104      \\
              & 1/area(95\%\, CL)          & 8.73 $\times 10^2$  
& 8.19 $\times 10^2$  \\
              & correlation                & -0.14      & -0.67     \\
\hline
\hline
standard      & $\sigma_{w_o}$ & 2.90 $\times 10^{-2}$ & 3.65 
$\times 10^{-2}$ \\
deviations    & $\sigma_{w_a}$ & 6.57 $\times 10^{-2}$ & 9.36 
$\times 10^{-2}$ \\
              & $\sigma_{\Omega_m}$ & 1.21 $\times 10^{-3}
$ & 1.18 $\times 10^{-3}$ \\
\hline
\end{tabular}
\vskip .25truecm
\caption{ Features of likelihood ellipses in the neighborhood of
$\Lambda$CDM, for the 5 bin case with $\ell = 2000$; the angle between
the major axis and the abscissa is dubbed $\theta$; the abscissa being
$w_0$ (former two cases) or $w_a.$ For the definition of the inverse
area and correlation see text.}
\label{tab2}
\end{table}


In particular, in Table \ref{tab2}, we give the inverse area of the
ellipses at 95\% CL~; when considered in the $w_0$--$w_a$ plane, it
coincides with the figure of merit
\begin{equation}
FOM = \frac{1}{{\rm area(95\%\, CL)}} = \frac{\sqrt{{\rm Det}|F_{ij}|}}{
4.61\, \pi}
\end{equation}
introduced in the report of the Dark Energy Task Force
\cite{albrecht}; the factor 4.61 yields 95\% CL (see eq.~(\ref{CL})).

The correlation coefficient or {\it Pearson correlation} measures the
correlation between two variables $X$, $Y$. In terms of Fisher Matrix
components, it reads
\begin{equation}
\rho (X,Y) = F^{-1}_{XY}/\sqrt{F^{-1}_{XX} \times F^{-1}_{YY}}~~.
\end{equation}

Table \ref{tab2} confirms that, when considering models different from
$\Lambda$CDM, non linear correction obtained through {\sc halofit} may
be misleading. This is true even when the fiducial model is
$\Lambda$CDM itself and we just consider mild deviations of $w$ from
$-1$. Let us point out, in particular, the $w_a$--$\Omega_m$ case;
here the parameter correlation increases by a factor $\sim 5$, when
passing from {\sc halofit} to simulations, because of a radical change
in the orientation of the ellipse, as the angle between its major axis
and the $w_a$ axis passes from -0.15$^o$ to -0.49$^o$ (see also figure
\ref{lcdm2000}).

As expected, the error on $\Omega_m$ estimate is not affected by the
passage from simulations to {\sc halofit}; in this case we are dealing
with $\Lambda$CDM models only. On the contrary, using {\sc halofit}
leads to underestimates of the errors on $w_0$ and $w_a$, by a
substantial 30--40$\, \%~.$ Let us however outline that the
substantial shift of the $\Omega_m$--$w_a$ correlation means that the
very estimate of the matter density parameter $\Omega_m$ is in
jeopardy, as soon as we admit mild deviations from pure $\Lambda$CDM,
so that the density of DE has a slight dependence on redshift.

\subsection{Confidence ellipses: M1 \& M3 models}

\begin{figure}
\begin{center}
\includegraphics[scale=0.69]{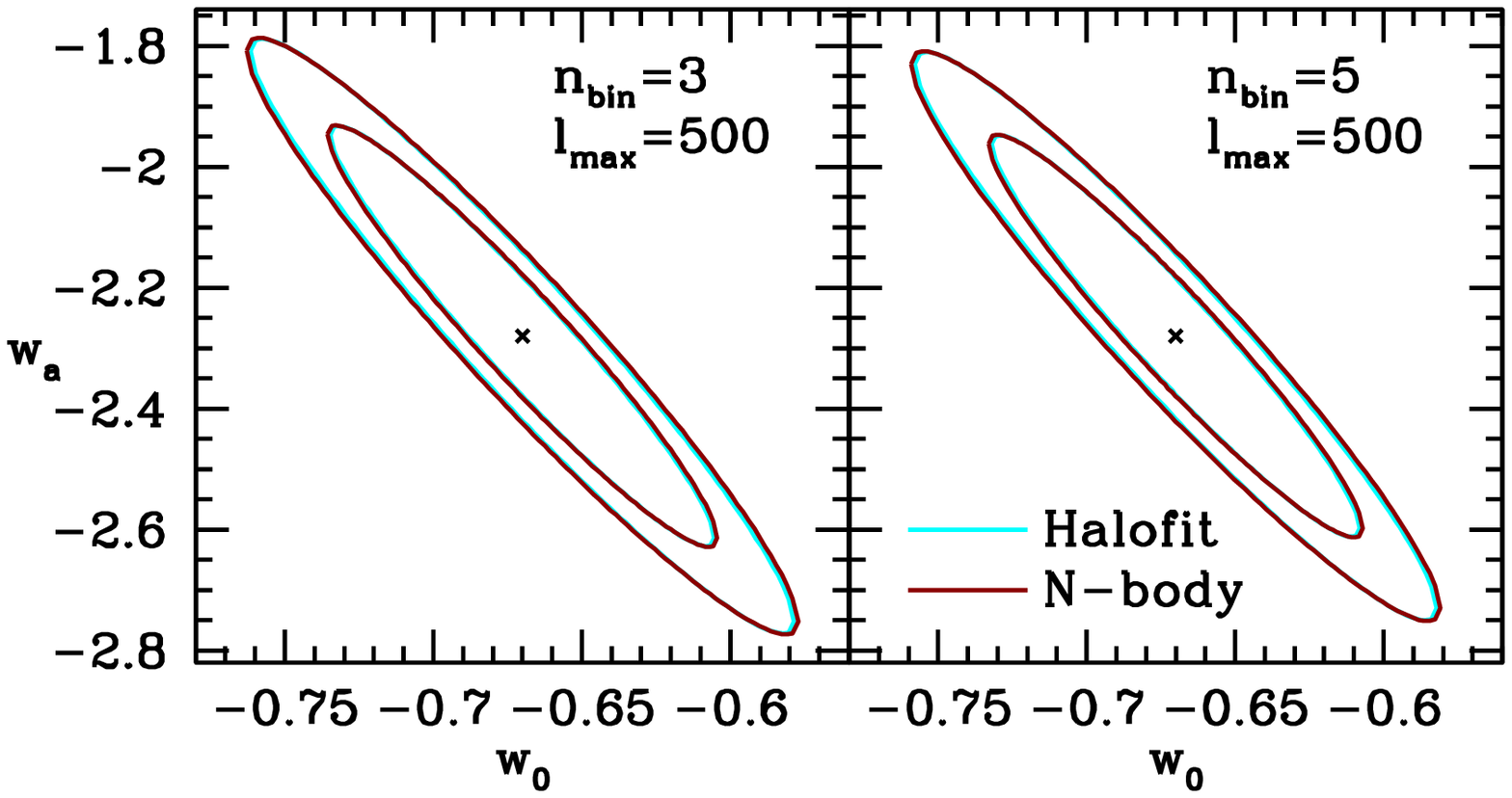}
\includegraphics[scale=0.69]{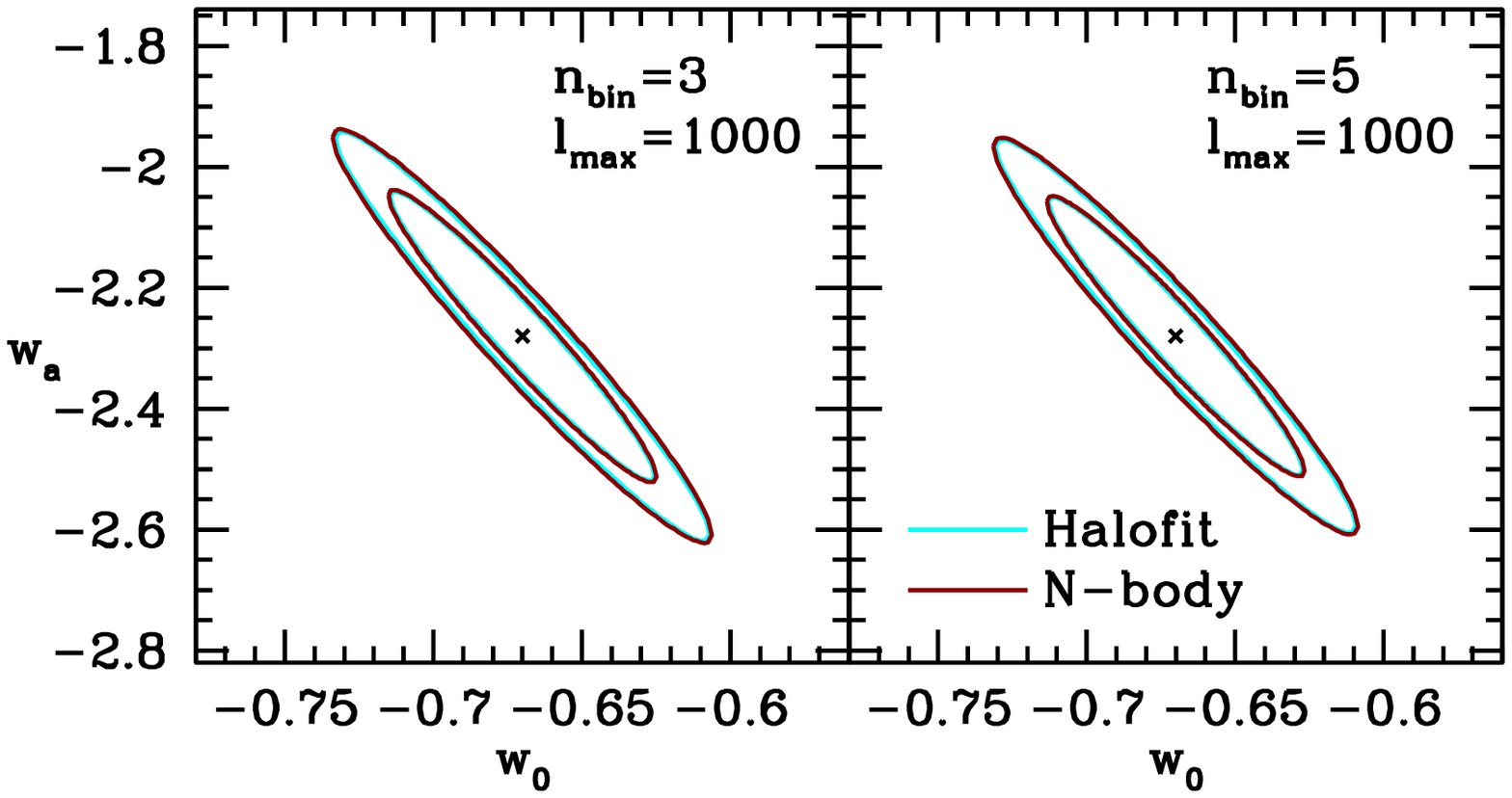}
\includegraphics[scale=0.69]{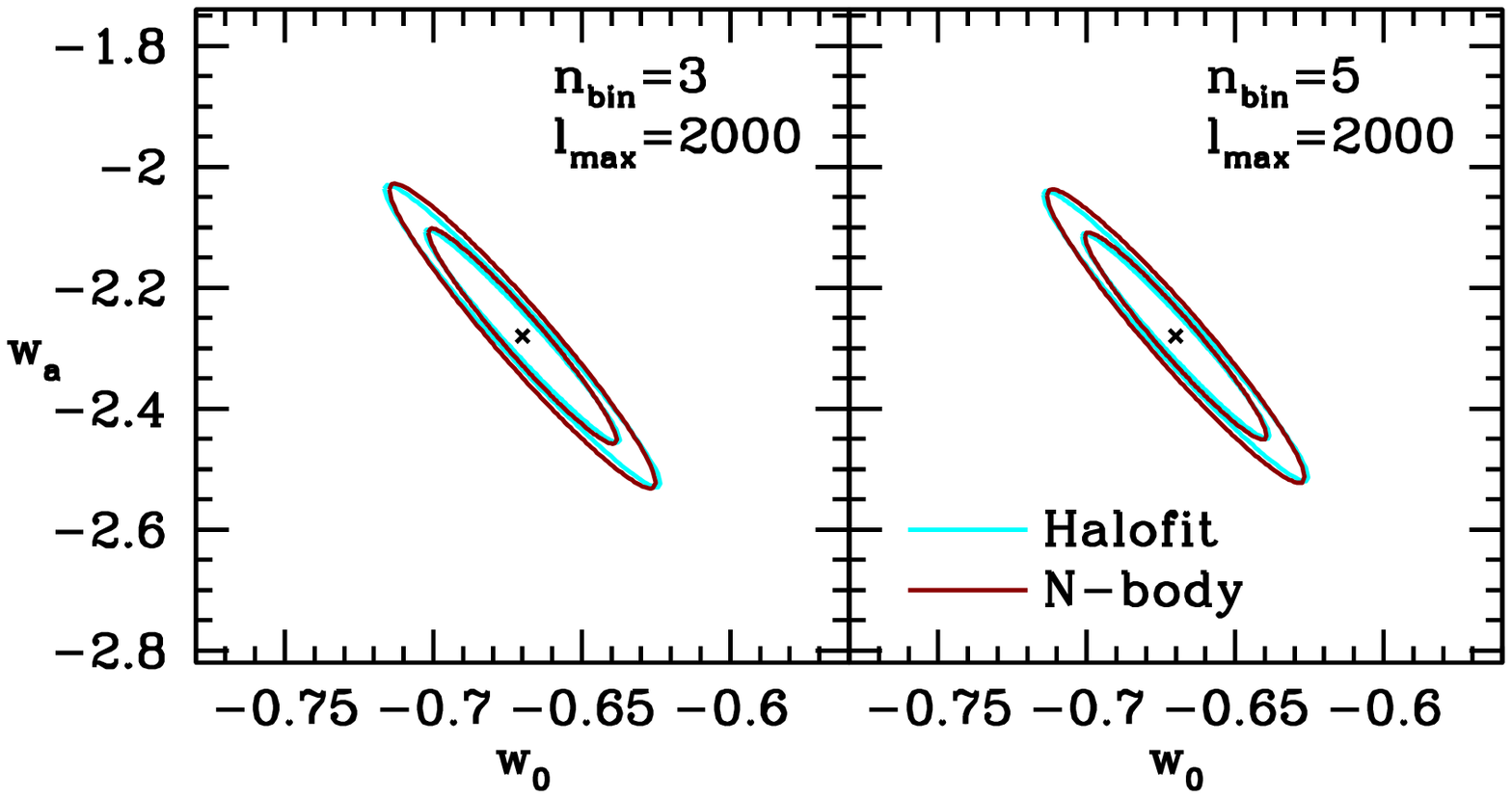}
\end{center}
\caption{Contours at 68\% and 95\% CL for the M1 model. {\sc halofit}
ellipses almost overlap with the $N$-body ones.}
\label{m1}

\end{figure}

\begin{figure}
\begin{center}
\includegraphics[scale=0.69]{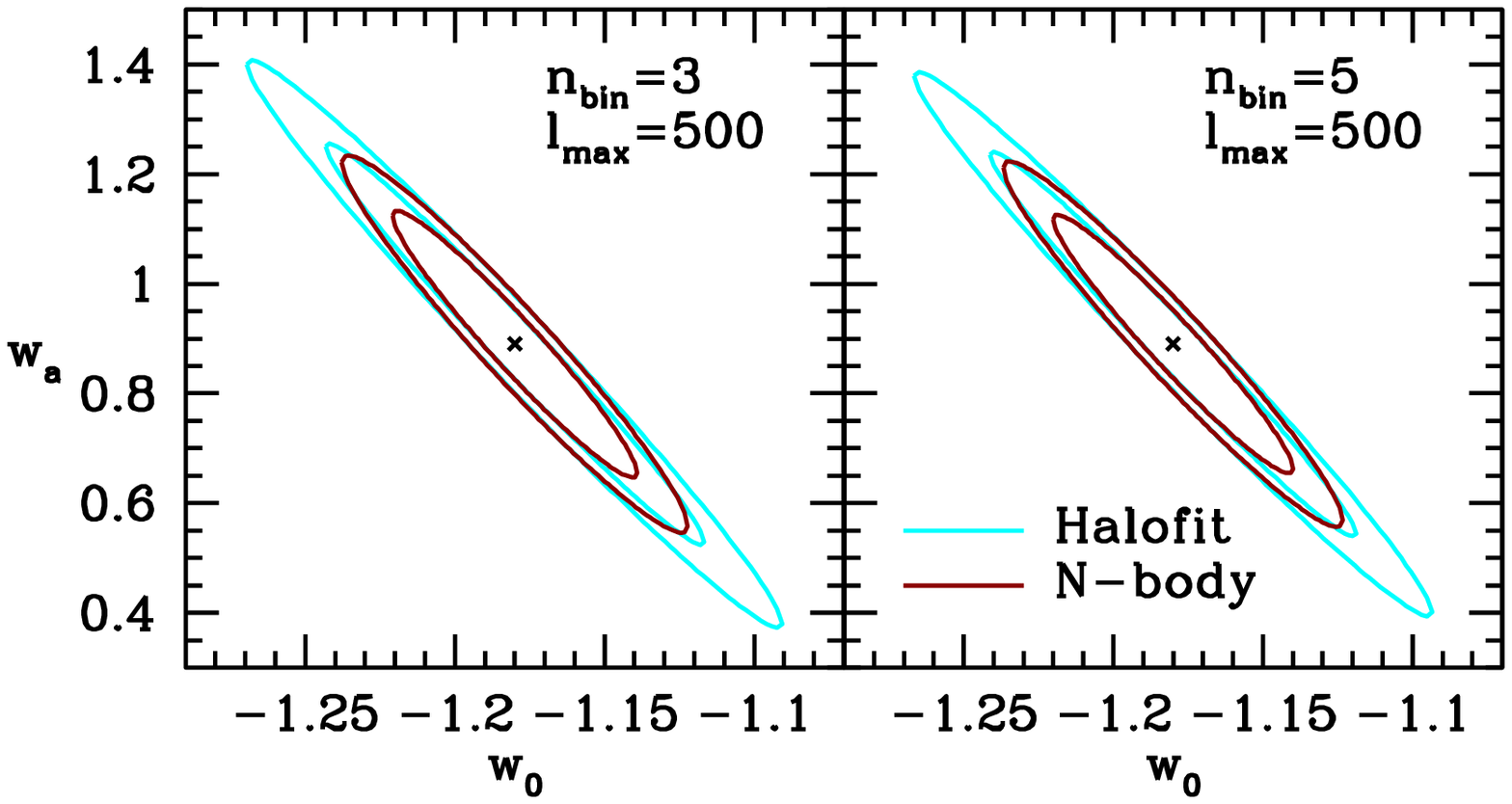}
\includegraphics[scale=0.69]{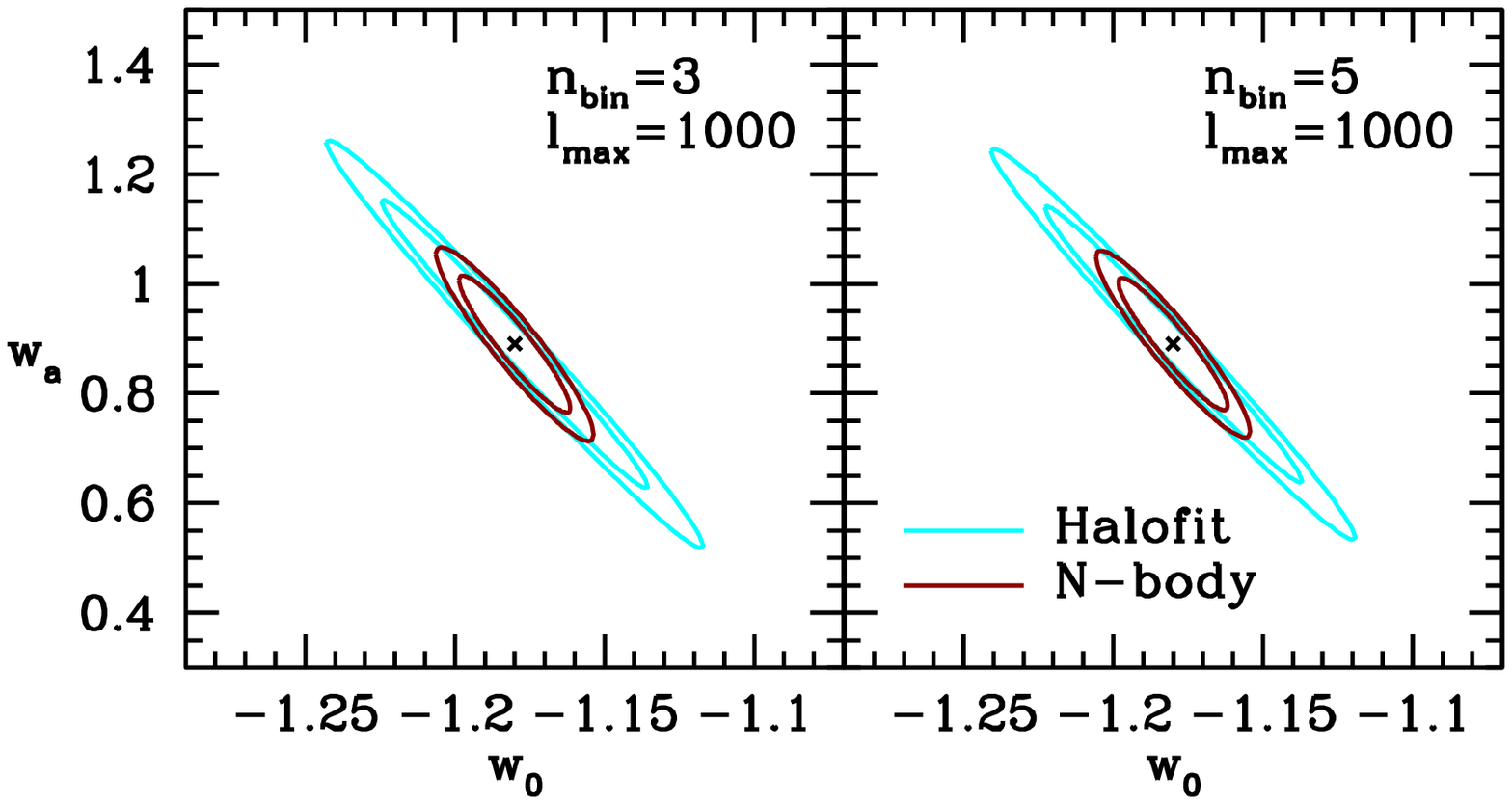}
\includegraphics[scale=0.69]{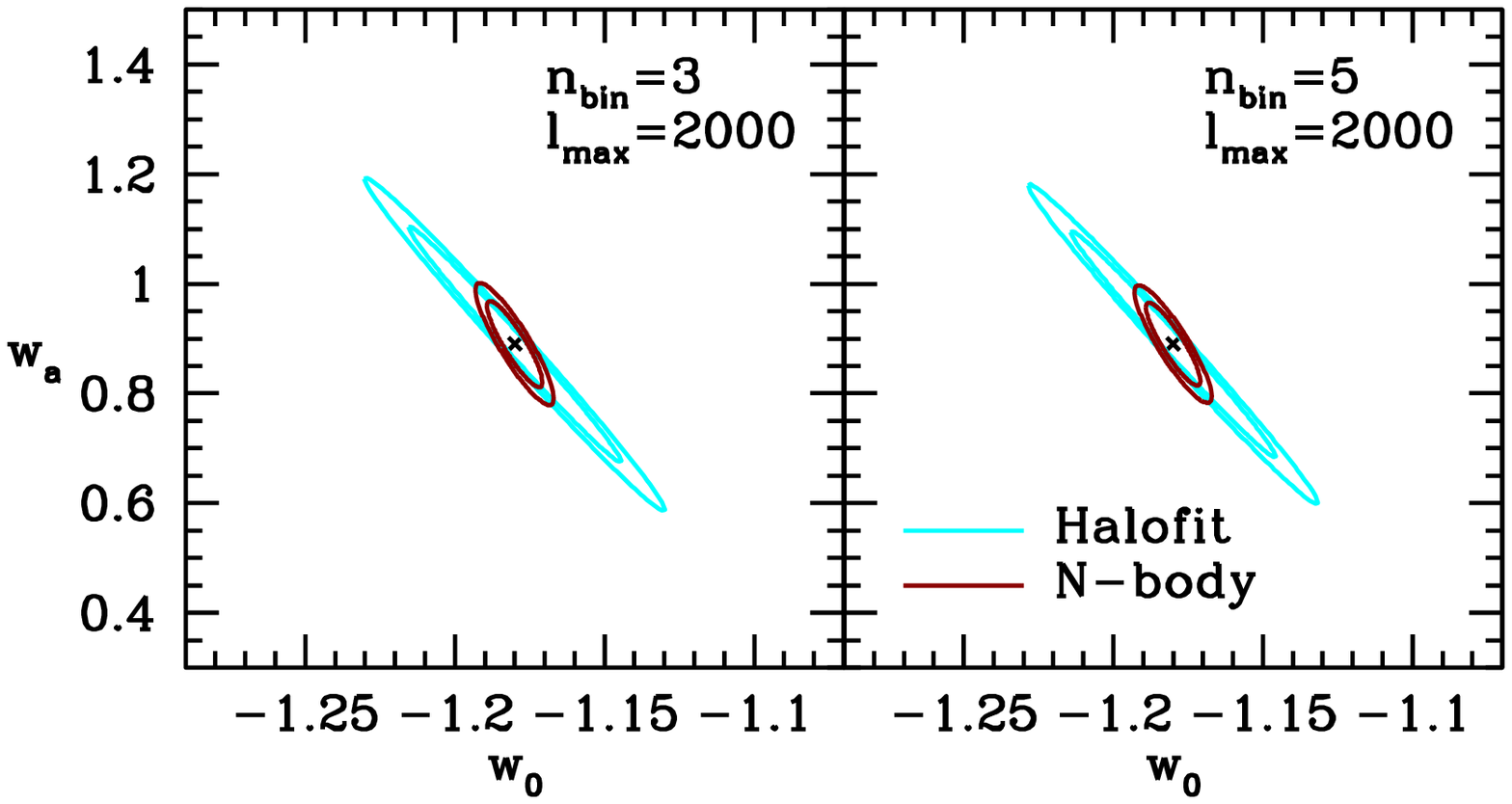}
\end{center}
\caption{Contours at 68\% and 95\% CL for the M3 model.}
\label{m3}
\end{figure}

Figures \ref{m1} and \ref{m3} then show the results in the
$w_0$--$w_a$ plane, when the fiducial models are M1 or M3. It is
evident that the two cases are quite different. This reflects the
behaviors of derivatives, whose ratios almost overlap in the M1 case,
while remaining different in the M3 case.
\begin{table}[t!]
\centering
\begin{tabular}{c c c c}
\hline
&&{\sc halofit}&sim. average\\
\hline
\hline
$w_o$--$w_a$  & $\theta$                   & $-80.6^o$  & $-83.9^o$ \\
 plane        & axial~ratio~(95\%\, CL) & 49.7       & 18.3      \\
              & 1/area(95\%\, CL)       & 181        & 505       \\
              & correlation                & -0.99      & -0.89     \\
\hline
\hline
standard      & $\sigma_{w_o}$ & 2.26 $\times 10^{-2}$ & 0.6 $\times 10^{-2}$ \\
deviations    & $\sigma_{w_a}$ & 0.136                 & 4.98 $\times 10^{-2}$\\
\hline
\end{tabular}
\caption{Features of likelihood ellipses when the fiducial model is
M3, for the 5 bin case with $\ell = 2000$; symbols as in Table
\protect\ref{tab2}.}
\label{tab3}
\end{table}

In the M1 case, we see just quite a mild shift. For $\ell = 500$ or
1000, and 3 or 5 bins, the discrepancy appears rather small, being
just slightly more evident for $\ell = 2000$. Let us however outline
that such mild discrepancies are $\cal O$ $(10 \, \%)$, compatible
with discrepancies attaining $\sim 20\, \%$ in the derivatives (see
figure \ref{derivative12}, left panel). It is also clear that
including larger $\ell$'s means integrating over slightly greater
discrepancies. The key issue, however, is that, until the two
derivative ratios proceed with similar trends, the only effect on
ellipses is a change in their area, while the orientation of their
axes does not change.

Let us now come to the M3 case, where the derivative ratios exhibit
radically different trends. The most immediate effect is that errors
estimated through {\sc halofit} exceed simulation errors by a
substantial factor. Discrepancies are already visible when only mildly
non--linear contributions are included and increase sharply when
exploiting more deeply non--linear spectral areas.

In Table \ref{tab3} we report quantitative information on the
deformation of likelihood ellipses on the 5-bin $\ell = 2000$ case.
The only elements which do not change much are the correlation and,
accordingly, the angle $\theta$. On the contrary, the ratio between
the axes of the ellipses and the FOM vary by a factor $\sim 2.5~.$
Standard deviations are also deeply affected.

Altogether, this is a case when estimates based on {\sc halofit} are
simply not trustworthy.

\section{Conclusions}
\label{sec:disc}
In the last decades, cosmological observations led to the most
important discoveries in fundamental physics. The dark cosmic
components are either a proof of physics beyond the standard model of
elementary interactions or a way to parametrize violations of General
Relativity. In the former case, the two dark components could be the
phenomenological description of a single cosmic fluid
\cite{BoMa}. Energy exchanges between such components could then be a
signal of such dark unity \cite{amend}. In turn, this opens a list of
possibilities which is embarrassingly long.

This {\it florilegium} of options is a clear consequence of the lack
of data. Tomographic weak lensing measures will clearly contribute to
provide more of them. This undoubtedly implies an important
experimental effort and there are scarce doubts that, accordingly,
data interpretation tools deserve to be updated.

In this work we outline the need to improve our capacity to predict
the spectra of cosmological models. When we restrict ourselves within
the frame of $\Lambda$CDM cosmologies, e.g. to perform derivatives in
respect to $\Omega_m$, the {\sc halofit} expressions still yield
acceptable outputs (see Figure \ref{derivative}).  
Improving their accuracy keeps however a need, namely for 
the parameter range consistent with most recent observations.

A first technical conclusion is that an inspection of the non--linear
part of spectra, through tomographic weak lensing surveys, greatly
improves their discriminatory power. This is also confirmed by the
effects of including an increasing number of $\ell$ harmonics in the
analysis: when {\sc halofit} and simulations yield discrepant results,
the difference between them increases when including greater and
greater $\ell$ values. On the contrary, with the instrumental features
assumed here, essentially corresponding to those of the {\sc Euclid}
\cite{euclid2} project, there seems to be no clear advantage in going
from 3 to 5 bins.

Our main conclusion is however that, going beyond a DE state equation
with $w \equiv -1$, direct model simulation remains the only efficient
way to obtain spectra. Some attempt to generalize {\sc halofit} to
constant $w \neq -1$, although important, do not hit the heart of the
problem, as we rather expect a variable $w(z)$. We should also be
aware of the severe bias on $w(z)$ detection which could follow the
assumption of constant $w$ \cite{bias}.

In this work we have tested the effects of using $\Lambda$CDM--suited
{\sc halofit} expressions, when the true cosmology is a simple dDE
model, straightforwardly consistent with WMAP7 data \cite{wmap7}.

We find that, even with $\Lambda$CDM as fiducial, when trying to
estimate the error on $w_0$ and $w_a$, we could be misled, if using
{\sc halofit} expressions. Not only they lead there to error
misestimates, but they can also imply a wrong correlation between the
estimate of $\Omega_m$ and, e.g., $w_a$. Henceforth, the very
$\Omega_m$ estimate is in jeopardy, if we do not treat accurately the
functional ensemble of state equations $w(z)$ about $w \equiv -1$.
This is, perhaps, the most striking result of this work.

We extended our test to the two dynamical DE models, differing from
$\Lambda$CDM just for $w(z)$, which are however within 95\% CL from
$\Lambda$CDM, on the basis of available data. Curiously enough, the
two models behave rather differently.

For one of them, the discrepancies between {\sc halofit} and
simulations, although visible in the derivatives, leave just $\cal O$
$(10\, \%)$ discrepancies in similarly oriented likelihood
ellipses. In this case error estimates performed through {\sc halofit}
turn out to be fair.

In the case of the other fiducial model, we meet an absolutely
different situation. First of all, using {\sc halofit} leads to error
overestimates by large factors (100 to 200$\%$ larger), to a wrong
figure of merit, and to other problems which can be summarized by
stating that {\sc halofit}--based results are wrong.

In the latter cases (non--$\Lambda$CDM) we have not yet tested the
intercorrelation between DE state equations and the density
parameter. This will be among our future tasks.

In spite of that, these results suggest a trend in the effectiveness
of {\sc halofit}, when non--$\Lambda$CDM cosmologies are involved. We
meet: (i) An apparent reliability of such expressions, when the
fiducial model is characterized by negative $w_a$. (ii) Significant
difficulties for the $\Lambda$CDM case ($w_a = 0$), where we expected
them to have their best performance. (iii) In the case of positive
$w_a$, finally, we conclude that {\sc halofit} expressions yield
misleading results.

We might then conjecture that an indicator of {\sc halofit} efficiency
is the sign of the derivative
\begin{equation}
w'(0) = \frac{dw}{dz }\, \, (z=0)~.
\end{equation}
When $w'(0) \geq 0 $, any use of {\sc halofit} expressions seems
dangerous. However, as we have been dealing only with DE state
equations which can be fully characterized at $z=0$, this conjecture
need to be tested for other cases.

Altogether, let us conclude that {\sc halofit}, in its present form,
is not adequate to deal with the degrees of freedom opening when
non--$\Lambda$CDM cosmologies are considered, as none can predict,
{\it a priori}, which can be the physical $w(z)$ behavior.

\acknowledgments

Most numerical simulations were performed on the PIA and PanStarrs2
clusters of the Max--Planck--Institut f\"ur Astronomie at the
Rechenzentrum in Garching. Spectral analysis was performed at
CINECA--Bologna under the CINECA--INAF agreement (2008--2010). The
financial supports of MIUR through the PRIN08 program and of the
Italian Space Agency (ASI), though the COFIS program are
acknowledged. L. A. acknowledges support from the DFG project TRR33
"The Dark Universe".

\newcommand{\APJ}{Astrophys.\ J.~}
\newcommand{\APJS}{Astrophys.\ J. \ Suppl.~}
\newcommand{\MNRAS}{Mon.\ Not.\ R.\ Aston.\ Soc.~} 
\newcommand{\PL}{Phys.\ Lett.~} 
\newcommand{\PR}{Phys.\ Rev.~} 
\newcommand{\PRL}{Phys.\ Rev.\ Lett.~} 
\newcommand{\AeA}{Astronom.\ Astrophys.~}

\end{document}